\documentstyle[12pt,aps]{revtex}

\input epsf.sty

\begin{document}
\title{$\sigma$-fields x Chiral Scalars in Nuclear Three Body Potentials}
\author{C. M. Maekawa\thanks{maekawa@ift.unesp.br}}
\address{Instituto de F\'\i sica Te\'orica, Univrsidade Estadual de S\~ao Paulo, Caixa Postal 875, 01405-900, S\~ao Paulo, SP, Brazil }
\author{M. R. Robilotta\thanks{robilotta@if.usp.br}}
\address{FINPE,Instituto de F\'\i sica, Universidade de S\~ao Paulo, Caixa Postal 66318, 05315-970, S\~ao Paulo, SP, Brazil}
\maketitle

\vspace{0.5cm} 
The scalar-isoscalar field of an effective chiral Lagrangian transforms differently in either linear or non-linear frameworks:
in the former case it is the counterpart of the pion whereas in the latter it is chiral invariant on its own.
We compare the predictions from these two models for nucleon interactions and find results which are identical for two-body and rather different for three-body potentials. 
Some qualitative features of three-body interactions are discussed.

\vspace{0.5cm}

\section{Introduction}

Since the fundamental work of Yukawa, the meson exchange theory has been
successfully applied to the description of the nucleon-nucleon force.
Nowadays it is well established that the one pion exchange potential (OPEP)
is dominant at large distances, yielding a strong force with spin and
isospin dependences.

The next layer of the interaction is associated with the exchange of two
pions. This system has the lightest mass beyond the OPEP and involves an
intermediate pion-nucleon ($\pi N$) amplitude in a kinematical region which
is not directly accessible to experiment. Therefore a proper theoretical
treatment of this component of the potential requires the use of information
based on dispersion relations, as pointed out long ago by Cottinghan and
VinMau \cite{Vin+}. The implementation of this idea led to the Paris
potential \cite{Lac+75}, which is rather successful in describing
experimental data.

An important feature of the $\pi N$ amplitude is chiral symmetry, as emphasized more than 25 years ago by Brown and Durso \cite{Bro+}.
Recently the application of chiral symmetry to the NN interaction has deserved much attention, initially in the restricted framework
of pion and nucleon degrees of freedom \cite{chi}, and nowadays it is fair to say that the theoretical formulation of this sector of the interaction is free of ambiguities. 
However, a Lagrangian containing only pions and nucleons cannot reproduce low-energy $\pi$N data and hence this very reliable part of the model has to be complemented \cite{Kol+,Rob+}.
The combination of chiral symmetry and experimental information about the intermediate $\pi N$ amplitude, based on dispersion relations, was brought into this problem in the last two years \cite{Rob+}, with a successful description of asymptotic $NN$ scattering data \cite{Bal+,Kai+}.

In contrast with the OPEP, the two-pion exchange potential (TPEP) produces an attractive interaction that depends little on spin and isospin, because it is dominated by the exchange of a scalar-isoscalar system. 
In many phenomenological potentials, the actual two-pion dynamics is simulated by the exchange of an effective scalar-isoscalar
meson, with a mass around 550 MeV. 
In the framework of chiral symmetry, this meson is usually identified with the counterpart of the pion in the linear sigma-model and
one obtains a prediction for its coupling constant to the nucleon.
The existence of a scalar-isoscalar meson is controversial, the main candidate being the f$_0$(400-1200), that may be present 
in $\pi\pi$ scattering \cite{PDG}. 
In the case of NN scattering, the TPEP is quite well accounted for by non-linear chiral dynamics, constrained by experimental information, and the scalar-isoscalar meson is unnecessary. 
According to the unwritten law of quantum mechanics, stating that processes which are not forbidden are compulsory, the exchange of two pions must be considered in any realistic description of the NN interaction and the use of the sigma field to simulate the actual TPEP gives rise to shortcomings. 
In particular, a scalar field with mass $m_s$ yields a central $NN$ potential proportional to $e^{-m_sr}/r$, whereas the spatial dependence of TPEP is closer to $\left( e^{-\mu r}/r\right) ^2$, $\mu $ being the pion mass \cite{Rob+,Kai+}. 
Moreover, the TPEP is proportional to $g^4$, where $g$ is the 
$\pi N$ coupling constant, whereas the exchange of a sigma is proportional
to $g^2$. In spite of all these problems, the use of an effective scalar
field may be useful in problems where simplicity is more important than
precision. As far as the former is concerned an effective scalar field
allows calculations at tree level, whereas the exchange of two pions
involves loop integrations.

If one is willing to use an effective scalar field in a calculation, there
are two possibilities at hand. The first one consists in employing the usual
sigma field of the linear model, which is the chiral partner of the pion.
The other possibility is to use a scalar field in the framework of
non-linear Lagrangians, which is chiral invariant and appears naturally when
the non-linear fields are obtained from the linear ones \cite{Wei}. These
two scalars fields couple differently to pions and nucleons and hence lead
to predictions which do not overlap for some specific processes. The main
purpose of this work is to explore these predictions in the case of
three-body forces. Our presentation is divided as follows: in sect. 2 we
compare linear and non-linear Lagrangians and in sect 3 we motivate the
physical predictions of the latter. In sect. 4 we derive the kernel of the
three-body force, which is fully developed in sect. 5. Finally, in sect 6 we
discuss qualitatively some features of the force associated with sigmas and
chiral scalars.

\section{Lagrangians}

In this section we introduce the Lagrangians describing both the sigma and
the chiral scalar. In the framework of linear dynamics the field $\sigma $
is the chiral partner of the pion field \mbox{\boldmath $\pi $ } since, for
an axial transformation $\delta ^A$, we have $\delta ^A\mbox{\boldmath$ \pi$
}\rightarrow \sigma $ and $\delta ^A\sigma \rightarrow -\mbox{\boldmath $\pi
$}$. In the case of non-linear dynamics, the pion field is represented by ${%
\mbox{\boldmath $\phi $}}$ and we also consider a new field $S$, which
corresponds to a generalization of the field $\sigma ^{\prime }$, introduced
long ago by Weinberg \cite{Wei}. The corresponding axial transformation are $%
\delta ^A{\mbox{\boldmath $\phi $}=}F({\mbox{\boldmath $\phi $}}^2) $, where 
$F({\mbox{\boldmath $\phi $}}^2)$ is a function of ${\mbox{\boldmath $\phi $}%
}^2$ and $\delta ^AS=0$. The last transformation implies that $S$ is a
chiral scalar.

The linear Lagrangian ${\cal L}_\sigma $ has the usual form 

\begin{eqnarray}
{\cal L}_\sigma &=&\frac 12\left(\partial _\mu \sigma \partial ^\mu\sigma 
+ \partial _\mu \mbox{\boldmath $\pi $}\cdot \partial ^\mu \mbox{\boldmath $\pi $} \right) 
-\frac{\mu ^2}2\left( \sigma ^2+\mbox{\boldmath $\pi $}%
^2\right) +\frac 12f_\pi \mu ^2\sigma +U\left( \sigma ^2+\mbox{\boldmath
$\pi $}^2\right)  \nonumber \\
&&+\bar Ni{\not \partial }N-g\bar N\left( \sigma +i{\mbox{\boldmath $\tau $}%
\cdot {\mbox{\boldmath$\pi$}} }\gamma _5\right) N  \label{Llin}
\end{eqnarray}

\noindent
where $N$ is the nucleon field that transforms linearly, $f_\pi $ is de pion
decay constant, $\mu $ is the pion mass, $g$ is the $\pi N$ coupling
constant and the term $U\left( \sigma ^2+\mbox{\boldmath $\pi $}^2\right) $
represents the self interactions of the mesonic fields. In the framework of
the linear sigma model the scalar fluctuations are associated with a field $%
\epsilon $, related to $\sigma $ by 

\begin{equation}
\sigma =f_\pi +\epsilon .
\end{equation}

Replacing this into the Lagrangian (\ref{Llin}), we obtain 

\begin{eqnarray}
{\cal L}_\sigma &=& \frac 12\left( \partial _\mu \epsilon \partial ^\mu \epsilon -m_\epsilon^2\epsilon ^2\right) 
+\frac 12\left( \partial _\mu \mbox{\boldmath $\pi $}\cdot \partial ^\mu \mbox{\boldmath $\pi$ }
-\mu ^2\mbox{\boldmath $\pi $}^2\right)
+U\left( \left( f_\pi +\epsilon \right) ^2+\mbox{\boldmath $\pi $}^2\right)  
\nonumber \\
&&+\bar Ni{\not \partial }N-g\bar N\left( f_\pi +\epsilon +i \mbox{\boldmath
$\tau $}\cdot \mbox{\boldmath $\pi$ }\gamma _5\right) N.  \label{L1}
\end{eqnarray}

In the non-linear approach, the Lagrangian ${\cal L}_S$ is written as 

\begin{equation}
{\cal L}_S=\left[ \frac 12\left( \partial _\mu S\partial ^\mu
S-m_s^2S^2\right) -U\left( S\right) \right] 
+\left[ \frac 12\left( \partial _\mu {\mbox{\boldmath $\phi $}}\cdot \partial ^\mu {\mbox{\boldmath $\phi $}}
+\partial _\mu f\partial ^\mu f\right) +f_\pi \mu ^2f\right] +{\cal L}_N,
\end{equation}

where $f$ corresponds to the function $f=\sqrt{f_\pi ^2-{\mbox{\boldmath
$\phi $}}^2}$. Formally, $U(S)$ describes the self interactions of the
scalar field and ${\cal L}_N$ represents both the nucleon sector and its
interactions with bosonic fields. This part of Lagrangian may be cast in
many different forms, two of which are widely employed in the literature. In
one of them, the pion-nucleon coupling is pseudo-vector (PV) and, in the
other, it is pseudo scalar (PS). For PV coupling, one has 

\begin{equation}
{\cal L}_N^{PV}=\bar \psi i\gamma _\mu D^\mu \psi -m\bar \psi \psi 
+\frac g{2m}\bar \psi \gamma _\mu \gamma _5{\mbox{\boldmath $\tau $}}\psi\cdot D^\mu {\mbox{\boldmath $\phi $}}
-g_sS\bar \psi \psi \;\;,  \label{Lpv}
\end{equation}

\noindent
where $\psi $ is the nucleon field that transforms non-linearly, $m$ is its
mass and $g_s$ represents the coupling of the nucleon to the scalar. In this
expression the pion and nucleon covariant derivatives are given by \cite
{Wei68} 

\begin{eqnarray}
D^\mu {\mbox{\boldmath $\phi $}} &=&\partial _\mu {\mbox{\boldmath $\phi $}}-
\frac 1{f+f_\pi }\partial ^\mu f{\mbox{\boldmath $\phi $}}, \\
D^\mu \psi &=&\left[ \partial ^\mu +i\frac 1{f_\pi \left( f+f_\pi \right) }
\frac{{\mbox{\boldmath $\tau $}}}2 \cdot \left( {\mbox{\boldmath $\phi $}}\times
\partial ^\mu {\mbox{\boldmath $\phi $}}\right) \right] \psi .
\end{eqnarray}

This Lagrangian has been used recently in the study of $\pi N$ form factors
in constituent quarks models \cite{Mae+97} and it is worth noting that its
last term describes a chiral invariant coupling between the scalar and the
nucleon.

In the case of $PS$ coupling, one has 

\begin{equation}
{\cal L}_N^{PS}=\bar Ni{\not \partial }N -g\bar N\left( f+i{\mbox{\boldmath
$\tau $}}\cdot {\mbox{\boldmath $\phi $}}\gamma _5\right) N 
-\frac{g_s}{f_\pi }S\bar N\left( f+i{\mbox{\boldmath $\tau $}}\cdot {\mbox{\boldmath
$\phi $}}\gamma _5\right) N,
\end{equation}
\label{33}

\noindent
where $N$ is the nucleon field that transforms linearly. The last term of
this expression has the same meaning as the corresponding one in eq. (\ref
{Lpv}).

On general grounds one knows that, in the framework of chiral symmetry,
results should not depend on the choice of ${\cal L}_N$ \cite{ref11,ref12}.
The equivalence between the PS and PV Lagrangians was verified explicitly in
the case of TPEP \cite{Rob+} and of pion-nucleon interactions in
constituent quark models \cite{Mae+97}. Because of this equivalence and of
the similarity with the linear Lagrangian, our discussions are set in PS
case. Thus our complete non-linear Lagrangian is written as 

\begin{eqnarray}
{\cal L}_S^{PS} &=&\frac 12\left( \partial _\mu S\partial ^\mu S-m_S^2S^2\right) 
+\frac 12\left( \partial _\mu {\mbox{\boldmath $\phi $}}\cdot \partial ^\mu {\mbox{\boldmath $\phi $}}
+\partial _\mu f\partial ^\mu
f\right) +f_\pi \mu ^2f+U\left( S\right)  \nonumber \\
&&+\bar Ni{\not \partial }N-\left( g+\frac{g_s}{f_\pi }S\right) \bar N\left(
f+i{\mbox{\boldmath $\tau $}}\cdot {\mbox{\boldmath $\phi $}}\gamma
_5\right) N.  \label{Lnl}
\end{eqnarray}

The interaction terms in eqs (\ref{Lnl}) and (\ref{L1}) have the same
scalar-nucleon and pion-nucleon vertices depicted in fig. 1a. However, the
non-linear approach also yields extra vertices, representing seagull
scalar-pion-nucleon and pion-pion-nucleon interactions, among others,
displayed in fig. 1b. These results indicate that, at tree level, both
Lagrangians produce the same $NN$ potential. On the other hand the extra
scalar-pion-nucleon vertex represents a genuine difference between the two
approaches and has consequences in processes such as pion absorption by a
two-nucleon system or three body forces.

\section{Effective Seagull}

The purpose of this section is to discuss the meaning of the effective
seagull interaction in terms of more basic processes. In order to do this,
we note that, in the framework of non-linear Lagrangians, the TPEP in the
pure nucleon sector is given by the five diagrams of fig. 2a \cite{Rob+},
which we may want to associate with an effective scalar exchange. When an
external pion is attached to these processes, we have the three
possibilities indicated in fig. 2b, the last one corresponding to an
effective seagull, whose dynamical meaning is given in fig 2c.

The nine diagrams associated with the effective seagull interaction can be
understood as arising from the product of the amplitudes $T^{ba}$and $%
P^{bca} $, where the former describes the pion-nucleon scattering $\pi
^a\left( k\right) N\left( p\right) \rightarrow \pi ^b\left( k^{\prime
}\right) N\left( p^{\prime }\right) $ and latter represents the contribution
from pion production, $\pi ^a\left( k\right) N\left( p\right) \rightarrow
\pi ^b\left( k^{\prime }\right) \pi ^c\left( q\right) N\left( p^{\prime
}\right) $, as indicated in fig 3. The composite amplitude is denoted by $A$
and given by 
\begin{equation}
A=-i\int \frac{d^4Q}{\left( 2\pi \right) ^4}T^{ba}P^{bca}\frac 1{k^2-\mu ^2}%
\frac 1{k^{\prime 2}-\mu ^2},
\end{equation}
where $Q=\frac 12\left( k-k^{\prime }\right) .$

Using the $\pi N$ vertices obtained from eq. (\ref{Lnl}), we have 
\begin{eqnarray}
P^{bca} &=&ig^3
\left\{ \left( \delta _{ab}\tau _c-\delta _{bc}\tau _a-\delta _{ac}\tau _b+i\epsilon _{bac}\right) \right. 
\nonumber\\
&&\times \bar u\left[ \frac{\not{k'}}{\left( p'+k'\right) ^2-m^2}
\; \gamma _5 \; \frac{\not{k}}{\left( p+k\right) ^2-m^2}\right] u 
\nonumber \\
&&+\delta _{ac}\tau _b \;
\bar u\left[ \gamma _5\frac{\not{k'}}{\left(p'+k'\right) ^2-m^2}\;\frac 1{gf_\pi }\right] u  
\nonumber\\
&&\left. +\delta _{bc}\tau _a \;
\bar u\left[ \frac 1{gf_\pi }\;\frac{\not{k}}{\left( p+k\right) ^2-m^2}\gamma _5\right] u\right\} .
\end{eqnarray}

The part of A corresponding to a scalar-isoscalar exchange is associated with the factor $\delta _{ab}$ in the $\pi N$ amplitude and hence it is proportional to  

\begin{eqnarray}
\delta _{ab}P^{bac} &=&ig^3\tau _c \;
\bar u\left[ \frac{\not{k'}}{\left(p'+k'\right) ^2-m^2}\; \gamma _5 \; \frac{\not{k}}{\left( p+k\right)^2-m^2}\right.
\nonumber\\
&&\left. -\gamma _5 \; \frac{\not{k'}}{\left( p'+k'\right) ^2-m^2}\;\frac 1{gf_\pi }
-\frac 1{gf_\pi }\;\frac{\not {k}}{\left(p+k\right) ^2-m^2}\; \gamma _5\right] u.
\end{eqnarray}

In order to identify the main contribution to this amplitude, we go to the
soft pion limit, by using $k=k^{\prime }=(k_0,0)$ and then take the limit $%
k_0\rightarrow 0$. We thus obtain 

\begin{equation}
\delta _{ab}P^{bca}\stackrel{soft}{\rightarrow }-\frac{ig^3}{4m^2}\tau_c \bar u \gamma _5u,
\end{equation}

which has the same structure as the effective scalar-nucleon seagull vertex predicted by the non-linear Lagrangian given by (\ref{Lnl}).

\section{Effective $\pi NN$ vertex}

As far as NN interactions are concerned both the linear and non-linear
Lagrangians, given by eqs. (\ref{L1}) and (\ref{Lnl}), yield potentials at
tree level, which are due to the exchanges of one pion and one scalar meson
associated with the diagrams of fig. 1a. In principle the scalar coupling
constants predicted by these Lagrangians could be different. However, we are
interested only in dynamical distinctions between the sigma and the chiral
scalar and, from now on, we set $g_s=g$ in order to force both models to
simulate exactly the same two-body potential.

On the other hand, these models produce different predictions for the
effective $\pi NN$ vertex and in this section we consider the form of this
interaction. The basic ingredients are the amplitudes $T_\sigma $ and $T_s$,
that describe the processes $\pi N\rightarrow \sigma N$ and $\pi
N\rightarrow SN$ respectively. The amplitude $T_\sigma $ is determined by
just the first two diagrams of fig. 4a whereas $T_s$ also includes the
seagull term. As the calculations of $T_\sigma $ and $T_s$ are quite
similar, we denote both amplitudes by $T$ and identify the seagull
contribution by a parameter $\lambda $, such that $T_\sigma =T\left( \lambda
=0\right) $ and $T_s=T\left( \lambda =1\right) $. At tree level, $T$ is
given by the diagrams depicted in fig 4a, and we have 

\begin{equation}
T\left( \lambda \right) =-ig^2\tau _c\bar u\left( {\mbox{\boldmath $p'$}}\right)
\left[ \frac{\not{p}_d+m}{p_d^2-m^2}\gamma _5+\gamma _5\frac{\not{p}_x+m}{p_x^2-m^2}
+\frac \lambda {gf_\pi }\gamma _5\right] u\left( {\mbox{\boldmath $p $}}\right) .  
\label{alambda0}
\end{equation}

with $p_d=p+k$, $p_x=p^{\prime }-k$. Using Dirac equation, we rewrite (\ref{alambda0}) as 

\begin{equation}
T\left( \lambda \right) =-ig^2\tau _c\bar u\left( {\mbox{\boldmath $p'$}}\right) 
\left[ \frac{\not{k}}{p_d^2-m^2}+\frac{\not{k}}{p_x^2-m^2}+\frac \lambda
{gf_\pi }\right] \gamma _5u\left( {\mbox{\boldmath $p $}}\right) .
\label{alambda}
\end{equation}

It is worth pointing out that, as shown in \cite{Mae+97}, this result does
not depend on our choice of the $PS$ coupling in non-linear Lagrangian and
it would be the same if the $PV$ scheme were adopted.

This amplitude contains positive frequency states, which do not enter in the
construction of the proper $\pi NN$ kernel. In order to isolate these
contributions, we write the nucleon propagator as 
\begin{equation}
\frac{\not p+m}{p^2-m^2}=\frac 1{2E}\left[ \frac 1{p^0-E}\sum_su^s\left( {%
\mbox{\boldmath $p $}}\right) \bar u^s\left( {\mbox{\boldmath $p $}}\right) +%
\frac 1{p^0+E}\sum_sv^s\left( -{\mbox{\boldmath $p $}}\right) \bar v^s\left(
-{\mbox{\boldmath $p $}}\right) \right] ,
\end{equation}
\noindent
where $E=\sqrt{{\mbox{\boldmath $p $}}^{\ 2}+m^2}$. Thus 
\begin{equation}
\left[ \frac{\not p+m}{p^2-m^2}\right] _{(+)}=\frac 1{2E}\left[ \frac{\not
p+m}{p^0-E}-\gamma ^0\right]
\end{equation}
and the positive energy contribution is 
\begin{eqnarray}
T_{\left( +\right) } &=&-ig^2\tau _c\bar u\left( {\mbox{\boldmath $p $}}%
^{\prime }\right) \left[ \frac{\not k}{2E_d\left( p_d^0-E_d\right) }+\frac{%
\not k}{2E_x\left( p_x^0-E_x\right) }\right.  \nonumber \\
&&\left. -\left( \frac 1{2E_d}-\frac 1{2E_x}\right) \gamma ^0\right] \gamma
_5u\left( {\mbox{\boldmath $p $}}\right) ,  \label{Tmais}
\end{eqnarray}

\noindent with $\tilde p_i=\left( E_i,\vec p_i\right) $ for $i=d,x$.

Subtracting (\ref{Tmais}) from (\ref{alambda}) we get $\bar T$, the
irreducible positive frequency amplitude 

\begin{eqnarray}
\bar T\left( \lambda \right) &=&-ig^2\tau _c\bar u\left( {\mbox{\boldmath $p
$}}^{\prime }\right) \left[ -\frac{\not k}{2E_d\left( p_d^0+E_d\right) }-%
\frac{\not k}{2E_x\left( p_x^0+E_x\right) }\right.  \label{Tbar}
\nonumber\\
&&+\left. \left( \frac 1{2E_d}-\frac 1{2E_x}\right) \gamma ^0+\frac \lambda {%
gf_\pi }\right] \gamma _5u\left( {\mbox{\boldmath $p $}}\right) .
\end{eqnarray}

This allows the proper kernel for the process $\pi N_jN_k$ as shown in fig 4b, to be written as 

\begin{eqnarray}
K\left( \lambda \right) &=&-ig^3\left\{ \tau _c\bar u\left( {\mbox{\boldmath
$p $}}^{\prime }\right) \left[ -\frac{\not k}{2E_d\left( p_d^0+E_d\right) }-%
\frac{\not k}{2E_x\left( p_x^0+E_x\right) }\right. \right.  \nonumber \\
&&+\left. \left. \left( \frac 1{2E_d}-\frac 1{2E_x}\right) \gamma ^0+\frac 
\lambda {gf_\pi }\right] \gamma _5u\left( {\mbox{\boldmath $p $}}\right)
\right\} ^{(j)}  \nonumber \\
&&\times \frac 1{q^2-m_s^2}\left[ \bar u\left( {\mbox{\boldmath $p $}}%
^{\prime }\right) u\left( {\mbox{\boldmath $p $}}\right) \right] ^{(k)},
\label{TvwpiNN}
\end{eqnarray}

where $q=p_2^{\prime }-p_2$.

\section{Three body force}

In this section we derive the three-nucleon force due to the simultaneous
exchanges of one pion and one effective scalar meson and the basic diagram
is shown in figure 5. It produces the following three-nucleon amplitude 

\begin{eqnarray}
T^{ijk}(\lambda ) &=&g^4{\mbox{\boldmath $\tau $}}^{(i)}\cdot {%
\mbox{\boldmath $\tau $}}^{(j)}\left[ \bar u\left( {\mbox{\boldmath $p $}}%
^{\prime }\right) \gamma _5u\left( {\mbox{\boldmath $p $}}\right) \right]
^{(i)}\frac 1{k^2-\mu ^2}  \nonumber \\
&&\times \left\{ \bar u\left( {\mbox{\boldmath $p $}}^{\prime }\right)
\left[ -\frac{\not k}{2E_d\left( p_d^0+E_d\right) }-\frac{\not k}{2E_x\left(
p_x^0+E_x\right) }\right. \right.  \nonumber \\
&&+\left. \left. \left( \frac 1{2E_d}-\frac 1{2E_x}\right) \gamma ^0+\frac 
\lambda {gf_\pi }\right] \gamma _5u\left( {\mbox{\boldmath $p $}}\right)
\right\} ^{(j)}  \nonumber \\
&&\times \frac 1{q^2-m_s^2}\left[ \bar u\left( {\mbox{\boldmath $p $}}%
^{\prime }\right) u\left( {\mbox{\boldmath $p $}}\right) \right] ^{(k)}.
\end{eqnarray}

In order to obtain the non-relativistic limit of $T^{ijk}(\lambda )$, denoted by $t^{ijk}(\lambda )$, we use 

\begin{eqnarray*}
&&\frac 1{2E_d}-\frac 1{2E_x}\sim \frac{{\mbox{\boldmath $p $}}^2}{m^3}, \\
&&\frac 1{2E_d\left( p_d^0+E_d\right) }+\frac 1{2E_x\left( p_x^0+E_x\right) }%
\sim \frac 1{2m^2},
\end{eqnarray*}
\begin{eqnarray*}
\bar u\left( {\mbox{\boldmath $p $}}^{\prime }\right) u\left( {%
\mbox{\boldmath $p $}}\right) &\rightarrow &I, \\
\bar u\left( {\mbox{\boldmath $p $}}^{\prime }\right) \gamma _5u\left( {%
\mbox{\boldmath $p $}}\right) &\rightarrow &\frac 1{2m}{\mbox{\boldmath
$\sigma $} }\cdot \left( {\mbox{\boldmath $p $}}-{\mbox{\boldmath $p $}}%
^{\prime }\right) , \\
\bar u\left( {\mbox{\boldmath $p $}}^{\prime }\right) \gamma ^0\gamma
_5u\left( {\mbox{\boldmath $p $}}\right) &\rightarrow &\frac 1{2m}{%
\mbox{\boldmath $\sigma $} }\cdot \left( {\mbox{\boldmath $p $}}+{%
\mbox{\boldmath $p $}}^{\prime }\right) , \\
\bar u\left( {\mbox{\boldmath $p $}}^{\prime }\right) \not k\gamma _5u\left( 
{\mbox{\boldmath $p $}}\right) &\rightarrow &-{\mbox{\boldmath $\sigma $} }%
\cdot {\mbox{\boldmath $k $},} \\
\frac 1{k^2-\mu ^2} &\cong &-\frac 1{{\mbox{\boldmath $k $}}^2+\mu ^2}, \\
\frac 1{q^2-m_s^2} &\cong &-\frac 1{{\mbox{\boldmath $q $}}^2+m_s^2},
\end{eqnarray*}

where the arrows indicate that the normalization of the spinors were also changed. 
Using these results and keeping only terms of the order ${\bf p}/m$, we get the non-relativistic amplitude 

\begin{eqnarray}
t^{ijk}(\lambda ) &=&\frac{g^4}{\left( 2m\right) ^2} {\mbox{\boldmath $\tau$}}^{(i)}\cdot {\mbox{\boldmath $\tau $}}^{(j)} \; {\mbox{\boldmath $\sigma $} }^{(i)}\cdot {\bf k }\; \frac 1{{\bf k}^2+\mu ^2}  \nonumber \\
&&\times {\mbox{\boldmath $\sigma $} }^{(j)}\cdot \left[ \left( \frac 1m -%
\frac \lambda{gf_\pi }\right) {\bf k } +\frac \lambda {gf_\pi }{\bf q}%
\right] \frac 1{{\bf q}^2+m_s^2},
\end{eqnarray}

\noindent
where ${\bf k }={\bf p}_i-{\bf {p'}}_i$and ${\bf q}= {\bf  p'} _k- {\bf p}_k$. 
The full non-relativistic amplitude $t_{3N}$ is given by the permutation over all indices $ijk.$

In momentum space the three body potential $W$ is defined by 
\begin{equation}
\left\langle {\mbox{\boldmath $p $}}_1^{\prime }{\mbox{\boldmath $p $}}%
_2^{\prime }{\mbox{\boldmath $p $}}_3^{\prime }\left| W\right| {%
\mbox{\boldmath $p $}}_1{\mbox{\boldmath $p $}}_2{\mbox{\boldmath $p $}}%
_3\right\rangle =-(2\pi )^3\delta ^3\left( {\mbox{\boldmath $p $}}_1^{\prime
}+{\mbox{\boldmath $p $}}_2^{\prime }+{\mbox{\boldmath $p $}}_3^{\prime }-{%
\mbox{\boldmath $p $}}_1-{\mbox{\boldmath $p $}}_2-{\mbox{\boldmath $p $}}%
_3\right) t_{3N}.  \label{W3N}
\end{equation}

We apply Fourier transform to (\ref{W3N}) in order to obtain the potential
in configuration space and we have 

\begin{equation}
\left\langle {\mbox{\boldmath $r $}}_1^{\prime }{\mbox{\boldmath $r $}}%
_2^{\prime }{\mbox{\boldmath $r $}}_3^{\prime }\left| W\right| {%
\mbox{\boldmath $r $}}_1{\mbox{\boldmath $r $}}_2{\mbox{\boldmath $r $}}%
_3\right\rangle =\delta \left( {\mbox{\boldmath $r $}}_1^{\prime }-{%
\mbox{\boldmath $r $}}_1\right) \delta \left( {\mbox{\boldmath $r $}}%
_2^{\prime }-{\mbox{\boldmath $r $}}_2\right) \delta \left( {\mbox{\boldmath
$r $}}_3^{\prime }-{\mbox{\boldmath $r $}}_3\right) W. 
\end{equation}

\noindent
and the component $W^{ijk}$ of the local potential $W$ is given by 

\begin{eqnarray}
W^{ijk} &=&\frac{g^4}{\left( 4\pi \right) ^2\left( 2m\right) ^2}\frac{\mu m_s%
}m{\mbox{\boldmath $\tau $}}^{\left( i\right) }\cdot {\mbox{\boldmath $\tau
$}}^{\left( j\right) }
\nonumber\\
&&\times \left[ \left( 1-\frac{\lambda m}{gf_\pi }\right) U\left(
m_sr_{kj}\right) {\mbox{\boldmath $\sigma $} }^{\left( i\right) }\cdot {%
\mbox{\boldmath $\nabla $ }}_{ji}{\mbox{\boldmath $\sigma $} }^{\left(
j\right) }\cdot {\mbox{\boldmath$\nabla $ 
}}_{ji}U\left( \mu r_{ji}\right) \right.  \nonumber \\
&& +\left. \frac{\lambda m}{gf_\pi }{\mbox{\boldmath $\sigma $} }^{(i)}\cdot 
{\mbox 
{\boldmath$\nabla$ }}_{ji}U\left( \mu r_{ji}\right) {\mbox{\boldmath$\sigma$
}}^{(j)} \cdot {\mbox{\boldmath $\nabla $}}_{kj}U\left( m_sr_{kj}\right)
\right] , 
\end{eqnarray}
where $U(\mu r)=\frac{\exp (-\mu r)}{\mu r}$ and $r_{kj}=r_k - r_j$.

We assume the approximate identity $\frac m{gf_\pi }\simeq 1$ and, similarly
to the procedure used in the two pion exchange three-nucleon force \cite
{Coe+}, we regularize the mesonic exchanges with the following dipole form
factor 
\begin{equation}
G\left( k^2\right) =\left( \frac{\Lambda ^2-\mu ^2}{\Lambda ^2-k^2}\right)
^2,
\end{equation}
where $\Lambda $ is a cut off parameter.

The regularization changes the Yukawa-type function into 
\begin{equation}
U\left( \mu r,\Lambda \right) =\frac{e^{-\mu r}}{\mu r}-\frac \Lambda \mu 
\frac{e^{-\Lambda r}}{\Lambda r}-\frac 12\frac \mu \Lambda \left( \frac{%
\Lambda ^2}{\mu ^2}-1\right) e^{-\Lambda r}
\end{equation}
and its derivatives are given by 
\begin{eqnarray}
\frac{\partial U\left( \mu r,\Lambda \right) }{\partial r_\alpha } &=&\mu 
\frac{r_\alpha }rU_1\left( \mu r,\Lambda \right) , \\
\frac{\partial ^2U\left( \mu r,\Lambda \right) }{\partial r_\alpha \partial
r_\beta } &=&\frac{\mu ^2}3\left[ \delta _{\alpha \beta }\left( U\left( \mu
r,\Lambda \right) -G\left( r\right) \right) \right.  \nonumber \\
&&\left. +\left( \frac{3r_\alpha r_\beta }{r^2}-\delta _{\alpha \beta
}\right) U_2\left( \mu r,\Lambda \right) \right] ,
\end{eqnarray}
where 
\begin{eqnarray}
U_1\left( \mu r,\Lambda \right) &=&-\left( \frac 1{\mu r}+1\right) \frac{%
e^{-\mu r}}{\mu r}+\frac{\Lambda ^2}{\mu ^2}\left( 1+\frac 1{\Lambda r}%
\right) \frac{e^{-\Lambda r}}{\Lambda r}  \nonumber \\
&&+\frac 12\left( \frac{\Lambda ^2}{\mu ^2}-1\right) e^{-\Lambda r}, \\
U_2\left( \mu r,\Lambda \right) &=&\left( 1+\frac 3{\mu r}+\frac 3{\mu ^2r^2}%
\right) \frac{e^{-\mu r}}{\mu r}-\frac{\Lambda ^3}{\mu ^3}\left( 1+\frac 3{%
\Lambda r}+\frac 3{\Lambda ^2r^2}\right) \frac{e^{-\Lambda r}}{\Lambda r} 
\nonumber \\
&&-\frac 12\frac \Lambda \mu \left( \frac{\Lambda ^2}{\mu ^2}-1\right)
\left( 1+\frac 1{\Lambda r}\right) e^{-\Lambda r}, \\
G\left( r\right) &=&\frac 12\frac \mu \Lambda \left( \frac{\Lambda ^2}{\mu ^2%
}-1\right) ^2e^{-\Lambda r}.
\end{eqnarray}

Substituting these results into the potential, we have 

\begin{eqnarray}
W^{ijk} &=&\frac 43\left( \frac{g\mu }{2m}\right) ^4\frac m{\left( 4\pi
\right) ^2}\frac{m_s}\mu {\mbox{\boldmath $\tau $}}^{(i)}\cdot {%
\mbox{\boldmath $\tau $}}^{(j)}  \nonumber \\
&&\times \left\{ \left( 1-\lambda \right) \left[ {\mbox{\boldmath $\sigma $} 
}^{(i)}\cdot {\mbox{\boldmath$\sigma $}}^{(j)}U\left( m_sr_{kj},\Lambda
\right) \left( U\left( \mu r_{ji},\Lambda \right) -G\left( r_{ji},\Lambda
\right) \right) \right. \right.  \nonumber \\
&&+\left. {\bf S}_{ij}\left( {\mbox{\boldmath $\hat r$}}_{ji},{%
\mbox{\boldmath $\hat r$}}_{ji}\right) U\left( m_sr_{kj},\Lambda \right)
U_2\left( \mu r_{ji},\Lambda \right) \right]  \nonumber \\
&& +\lambda \frac{m_s}\mu 
\left[ {\bf S}_{ij}\left( {\mbox{\boldmath $\hat r$}}_{ji},{\mbox{\boldmath $\hat r$}}_{kj}\right)
+{\mbox{\boldmath $\hat r$}}_{ji}\cdot {\mbox{\boldmath $\hat r$}}_{kj} \;
{\mbox{\boldmath$\sigma $} }^{(i)}\cdot { \mbox{\boldmath$\sigma $ } }^{(j)} \right] 
\nonumber\\
&& \left. \times U_1\left(\mu r_{ji},\Lambda \right) U_1\left( m_sr_{kj},\Lambda \right) 
\right\} ,
\end{eqnarray}

where 
\begin{equation}
{\bf S}_{ij}\left( {\mbox{\boldmath $ \hat r$}}_{ji},{\mbox{\boldmath$ \hat r$}}_{kj}\right) 
=3{\mbox{\boldmath $\sigma $} }^{(i)}\cdot {\mbox{\boldmath
$\hat r$}}_{ji}{\mbox{\boldmath $\sigma $} }^{(j)}\cdot {\mbox{\boldmath
$\hat r$}}_{kj}-{\bf \hat r}_{ji}\cdot {\mbox{\boldmath $\hat r$}}_{kj}{%
\mbox{\boldmath $\sigma $} }^{(i)}\cdot \mbox{\boldmath $\sigma $}^{(j)}.
\end{equation}

\section{Discussion}

Initially, we discuss the role of chiral symmetry in the results of the previous section. 
Inspecting eq.(25), one notes that all the terms of the potential contain two gradients, reflecting the fact that they come from eq.(22),
 which is a uniform second order polynomial in meson momenta, as expected from a calculation based on chiral 
symmetry.
This feature of the problem is independent of $\lambda$. 
On the other hand, the detailed form of the potential is quite sensitive to this parameter.
The value $\lambda=0$ corresponds to a scalar field, denoted by $\sigma$, which is the chiral partner of the pion.
The value $\lambda=1$, in turn, indicates processes based on a chiral invariant field S.
It is worth pointing out that results with $\lambda=1$ are, as they should, independent of the representation adopted for the pion
field and, in particular, of the use of either PS or PV pion-nucleon couplings.
So the parameter $\lambda$ describes dynamical processes which are genuinely different. 

At tree level, these effects cannot be distinguished in NN interactions, but they manifest themselves in the reactions 
NN $\rightarrow\pi$NN, $\pi$d$\leftrightarrow$NN, in axial form factors of nuclei and in three-body forces.
In the present work we have considered only the last kind of application.

Our results with $\lambda=1$ coincide with that presented by Coon, Pe\~na and Riska \cite{CPR}. 
These authors also obtained a contact interaction, employing a PV pion-nucleon coupling and a scalar-meson coupled to nucleons.
In the framework of chiral symmetry, this combination means that they have tacitly used a Lagrangian equivalent of our eq.(5) and
hence their $\sigma$ field corresponds to our S.

In this work we are concerned mainly with the differences between the cases $\lambda=0$ and $\lambda=1$ for three-body
forces.
The full exploration of this aspect of the problem would require a precise calculation of trinucleon observables,
 especially the binding energy.
In order to produce a preliminary qualitative indication of the role of the force considered in this work, 
we evaluate the expectation value $\left\langle S\right| W\left| S\right\rangle $, where $\left| S\right\rangle$ 
represents the principal S-state of the trinucleon, which is the basic component of its ground state wave function.

This state is written as \cite{Coe+} 

\begin{equation}
\left| S\right\rangle =S\left( {\mbox{\boldmath $\hat r$}}_{ji}\right)
\Gamma _{\frac 12t}^{\frac 12m}\left( {\sf a}\right) 
\end{equation}

\noindent
where $S\left( {\mbox{\boldmath $r $}}_{ji}\right) $ is the spatial component and 
$\Gamma _{\frac 12t}^{\frac 12m}\left( {\sf a}\right) $ is the antisymmetric spin-isospin wave function with $z$-components $m$ and $t$ respectively.

The action of the tensor operators $S_{ij}$ over $\left| S\right\rangle $
results in states with orbital angular momentum different from zero. Using 

\begin{equation}
\left[ \Gamma _{\frac 12t}^{\frac 12m}\left( {\sf a}\right) \right]
^{\dagger }{\mbox{\boldmath $\tau $}}^{(i)}\cdot {\mbox{\boldmath $\tau $}}%
^{(j)}\;{\mbox{\boldmath $\sigma $} }^{(i)}\cdot {\mbox {\boldmath $\sigma$ }%
}^{(j)}\Gamma _{\frac 12t}^{\frac 12m}\left( {\sf a}\right) =-3
\end{equation}
we obtain 
\begin{eqnarray}
\left[ \Gamma _{\frac 12t}^{\frac 12m}\left( {\sf a}\right) \right]
^{\dagger }W^{ijk}\Gamma _{\frac 12t}^{\frac 12m}\left( {\sf a}\right)
&=&-4\left( \frac{g\mu }{2m}\right) ^4\frac m{\left( 4\pi \right) ^2}\frac{%
m_s}\mu  \nonumber \\
&&\times \left\{ \left( 1-\lambda \right) U\left( m_sr_{kj},\Lambda \right)
\left[ U\left( \mu r_{ji},\Lambda \right) -G\left( \mu r_{ji},\Lambda
\right) \right] \right.  \nonumber \\
&&+\left. \cos \theta _j\lambda \frac{m_s}\mu U_1\left( \mu r_{ji},\Lambda
\right) U_1\left( m_sr_{kj},\Lambda \right) \right\}
\end{eqnarray}

\noindent
where $\cos \theta _j={\mbox{\boldmath $\hat r$}}_{ji}\cdot {\mbox{\boldmath$\hat r$}}_{kj}$. 
Using the results of \cite{Coe+} and neglecting the very short range function 
$G\left( \mu r_{ji},\Lambda \right) $ \cite{RI}, we have 

\begin{eqnarray}
\left[ \Gamma _{\frac 12t}^{\frac 12m}\left( {\sf a}\right) \right]
^{\dagger }W^{ijk}\Gamma _{\frac 12t}^{\frac 12m}\left( {\sf a}\right)
&=&-\left\{ \left( 1-\lambda \right) C_\sigma U\left( m_sr_{kj},\Lambda
\right) U\left( \mu r_{ji},\Lambda \right) \right.  \nonumber \\
&&+\left. \lambda C_S\cos \theta _jU_1\left( \mu r_{ji},\Lambda \right)
U_1\left( m_sr_{kj},\Lambda \right) \right\} ,
\end{eqnarray}

\noindent
where $C_\sigma $ and $C_S$ are the strength coefficients of sigma and
chiral scalar respectively, given by

\begin{eqnarray*}
C_\sigma &=&4\left( \frac{g\mu }{2m}\right) ^4\frac 1{\left( 4\pi \right) ^2}%
\frac{m_s}\mu m, \\
C_S &=&4\left( \frac{g\mu }{2m}\right) ^4\frac 1{\left( 4\pi \right) ^2}%
\left( \frac{m_s}\mu \right) ^2m.
\end{eqnarray*}

\noindent
Choosing $m_s=550MeV$ and $g=13.5$, the strength coefficients become $%
C_\sigma =96MeV$ and $C_S=378MeV$.

In fig. 6 we show equipotential plots for the choices $\lambda =0$ (graph $a$) and $\lambda =1$ (graph $b$), constructed by keeping two nucleons $1\;fm$ apart and varying the position of the third one. 
As the plots are symmetric under rotation around the x-axes the specification of a single quadrant describes the spatial energy distribution. 
Inspecting this figure one learns that the predictions from both models are rather different, indicating that the effective seagull is very important. 
In the linear approach, the interaction is attractive over a wide region, whereas the non-linear scalar produces a repulsion around the triangular configuration and these differences should show up in observables. 

One is aware that a study based on just S trinucleon waves can provide only rough indications, since it is well known that 
D waves do play an important role in trinucleons.
Nevertheless, in the absence of a detailed study, we may assume that the trends associated with our equipotential plots would 
reflect in the binding energy.
This assumption is supported by the results of ref.{\cite{CPR}} where a three-body force given by our eq.(22) with 
$\lambda=1$ was shown to produce a decrease of the binding energy which is rather welcome.

As a final comment, it is important to point out that the discussion presented in section 3 makes us to be biased towards the chiral scalar, but final conclusions must wait until a complete evaluation of the diagrams of fig 2c, which is now in progress.


\newpage
\vspace{4cm}The work of one of us (C.M.M.) was supported by FAPESP,
Brazilian Agency.




\newpage
{\large {\bf Figure Captions}} 
\vspace{2cm}

\epsfbox{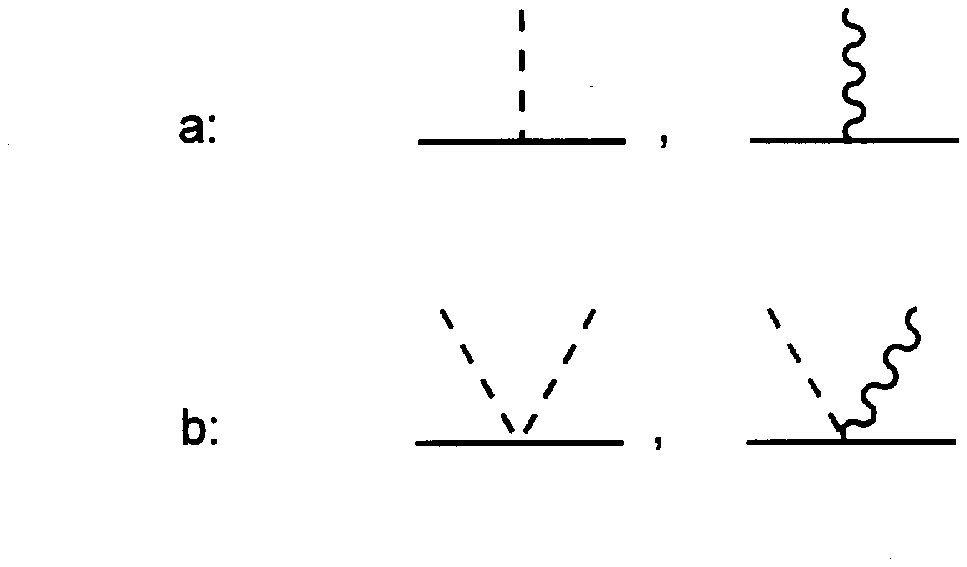}
Fig.1.Interactions of nucleons (full lines), pions (broken lines) and
scalar-isoscalar mesons (wavy lines) present in both linear and non-linear
models (a) and only in the latter (b).

\newpage
\epsfxsize = 17 cm
\epsfbox{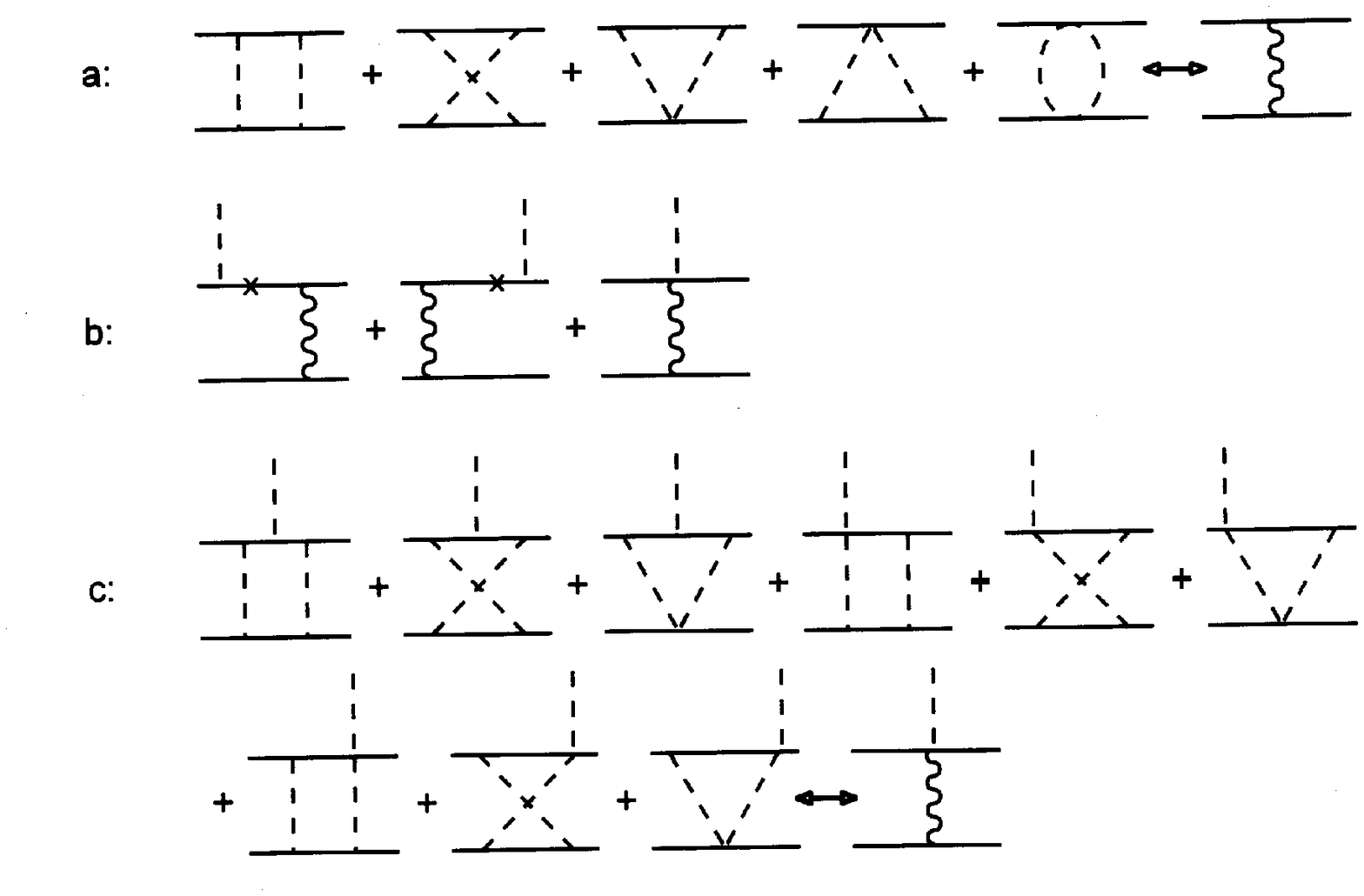}
Fig.2. Dynamical content of the effective scalar-isoscalar field in (a)
nucleon-nucleon interactions, (b) the $\pi NN$ kernel and (c) the seagull
term; conventions are the same as in fig.1 and the crosses in nucleon
propagators indicate that they do not contain positive frequency components.

\newpage
\epsfbox{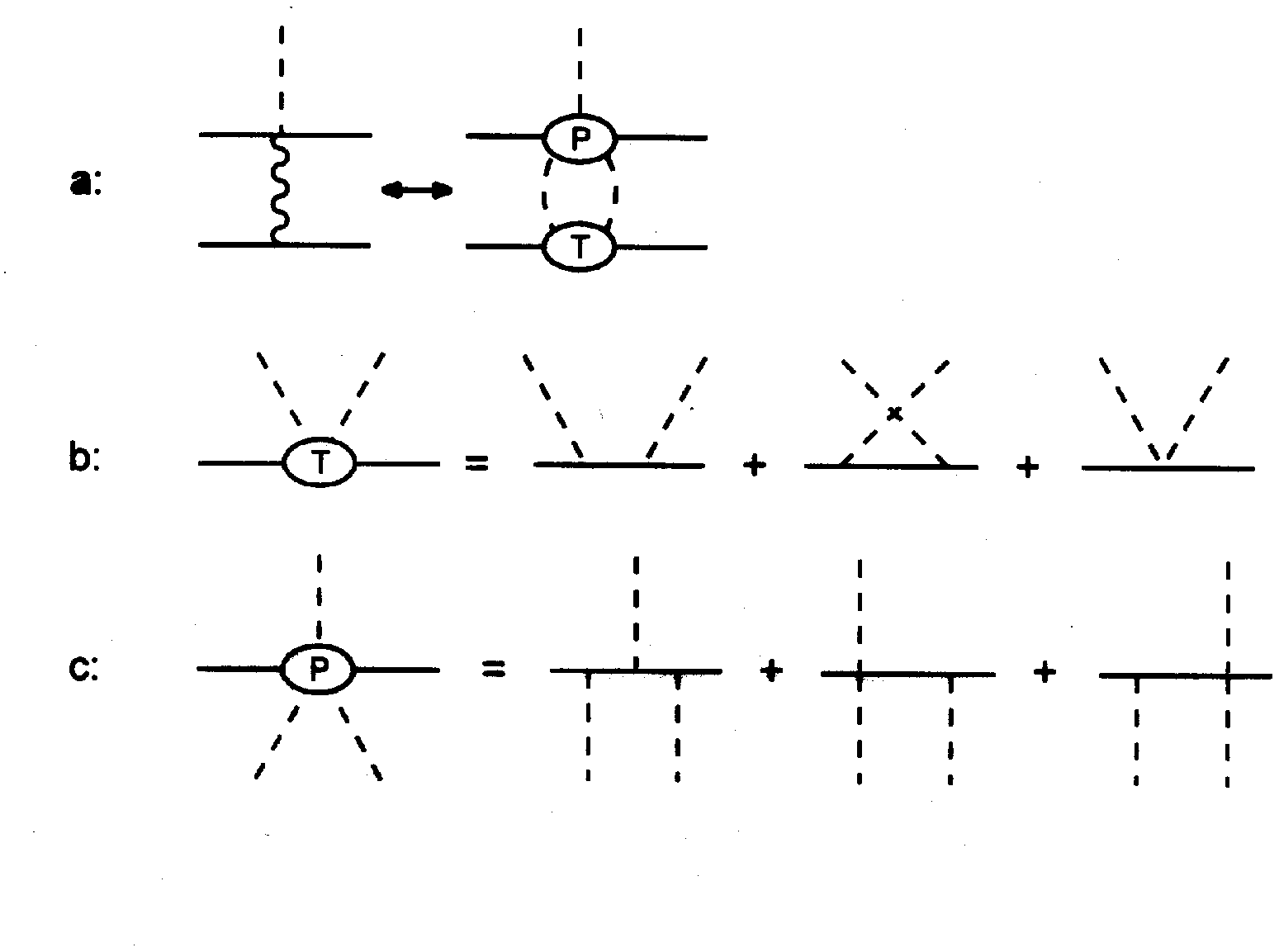}
Fig.3. The effective seagull (a) as composed by the $\pi N\rightarrow \pi N$
(b) and $\pi N\rightarrow \pi \pi N$ (c) amplitudes.

\newpage
\epsfbox{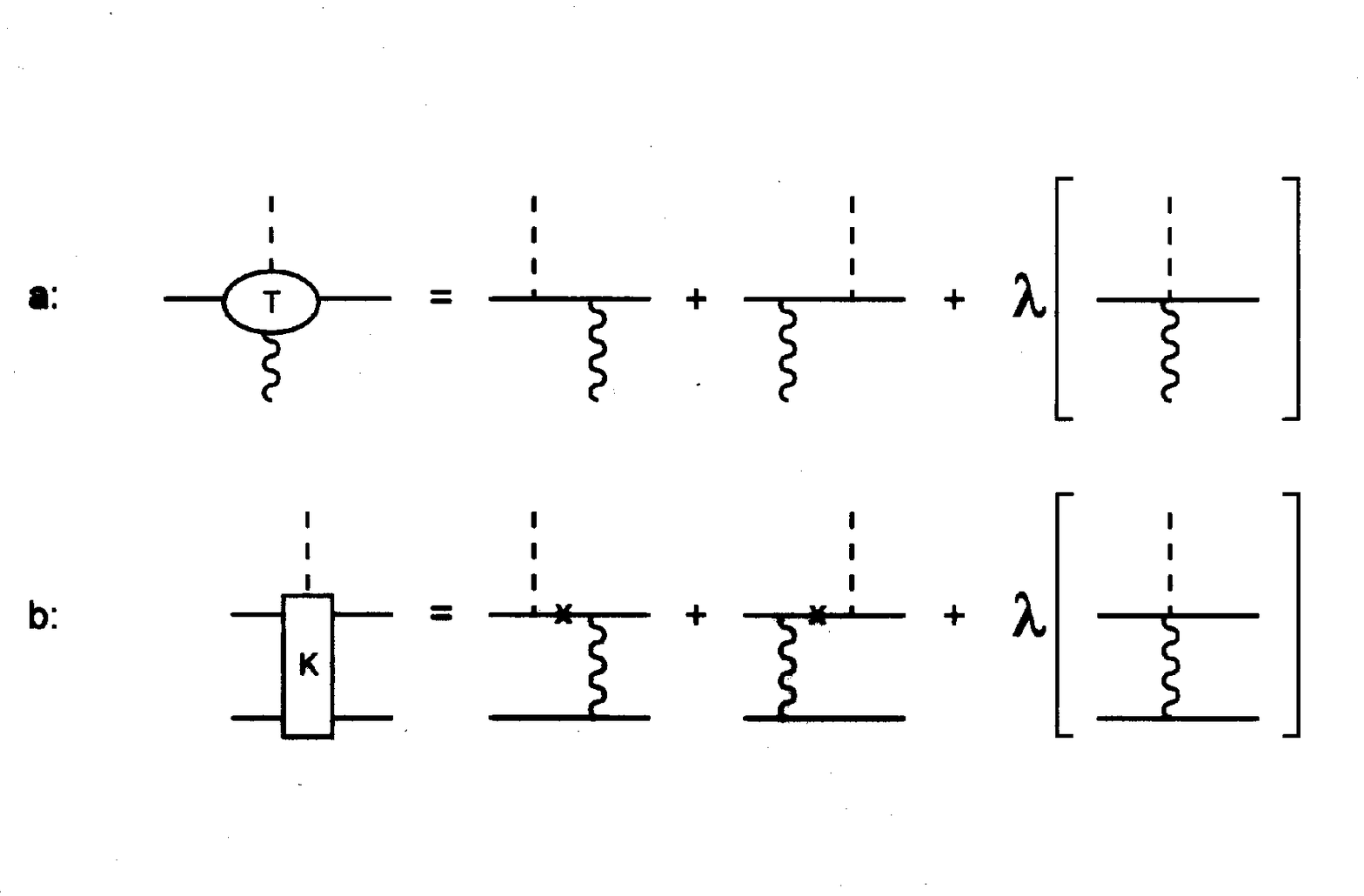}
Fig.4. (a) Amplitudes for the processes $\pi N\rightarrow \sigma N\;(\lambda
=0)$ and $\pi N\rightarrow SN\;(\lambda =1)$; (b) the corresponding $\pi NN$
kernel.

\newpage
\epsfbox{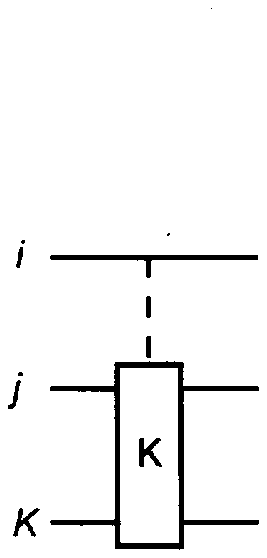}
Fig.5. Basic diagram for the three-body force.

\newpage
\epsfxsize=10 cm
\epsfbox{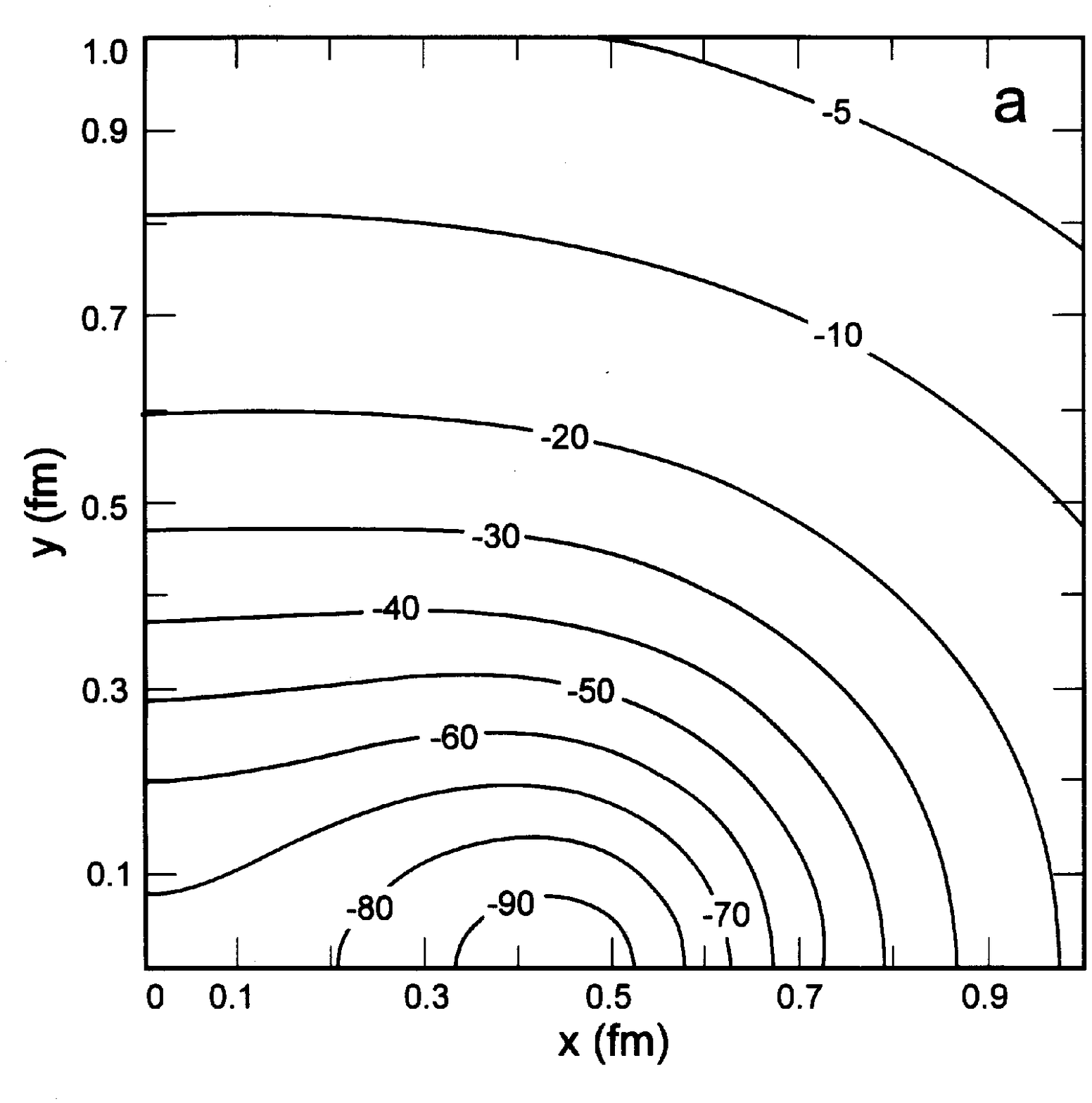}
\epsfxsize=10 cm
\epsfbox{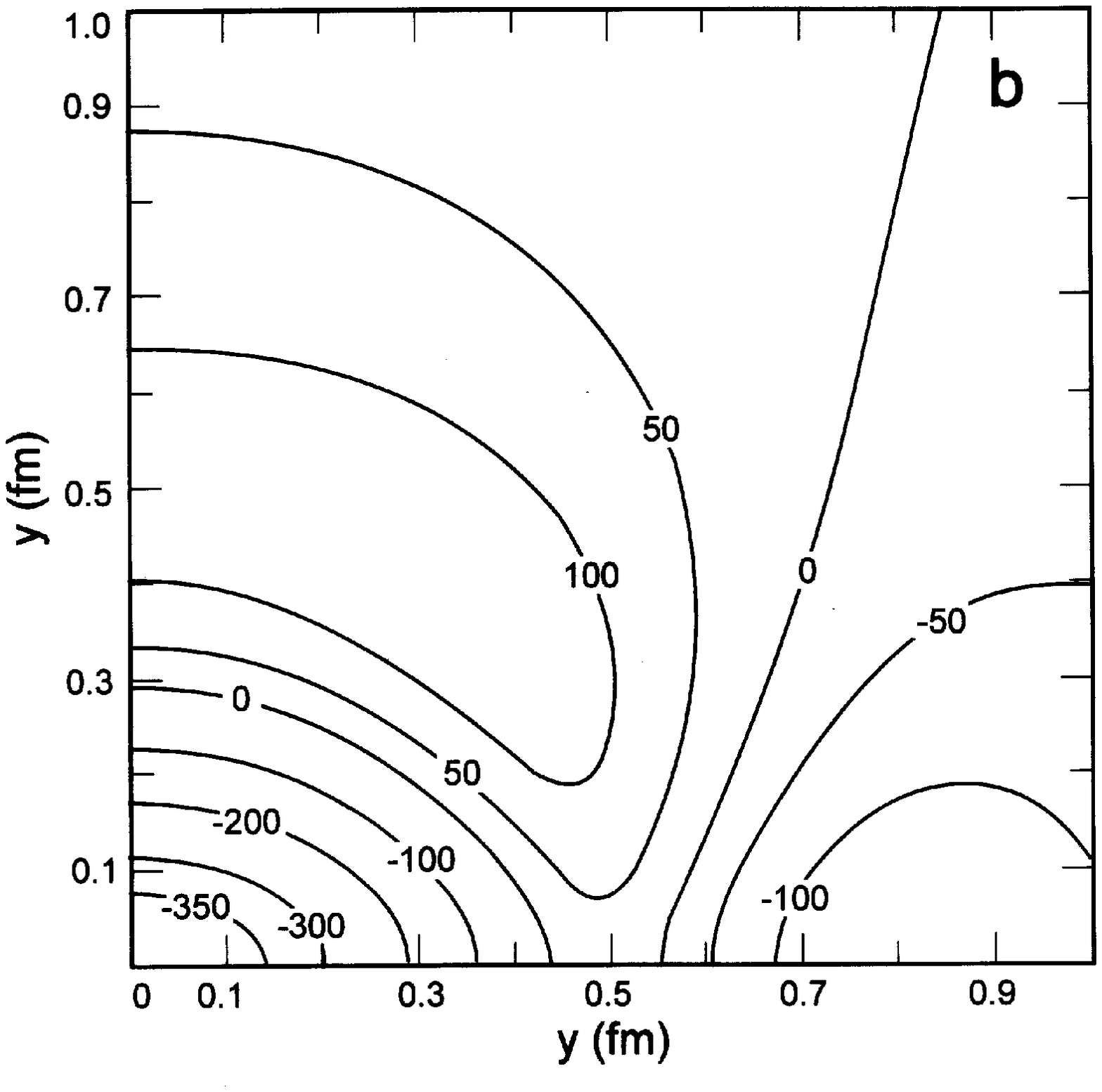}
Fig.6 Equipotential plots for the expectation value of the three body force
in the trinucleon ground state, calculated using eq (36) with $\Lambda
=1.5GeV$ for the linear (a) and non linear (b) models. One of the nucleons
is fixed at $x=0.5$ $fm$, another at $x=-0.5$ $fm$ and the position of the
third one is varied.

\end{document}